\documentclass[twocolumn]{aastex7}

\usepackage{amsmath}
\usepackage{hyperref}

\usepackage{xspace}

\newcommand{\de}{{\rm d}}
\newcommand{\feh}{{\rm [Fe/H]}}

\usepackage[range-phrase=-, range-units=single]{siunitx}

\hypersetup{pdfauthor={Name}}
\begin{document}

\title{The Kinematic Age of 3I/ATLAS and its Implications for Early Planet Formation}

 \author[0000-0002-0140-4475]{Aster G. Taylor}
 \altaffiliation{Fannie and John Hertz Foundation Fellow}
\affiliation{Dept. of Astronomy, University of Michigan, Ann Arbor, MI 48109, USA}
\email[show]{agtaylor@umich.edu}

\author[0000-0002-0726-6480]{Darryl Z. Seligman}
\altaffiliation{NSF Astronomy and Astrophysics Postdoctoral Fellow}
\affiliation{Department of Physics and Astronomy, Michigan State University, East Lansing, MI 48824, USA}
\email{dzs@msu.edu}

\begin{abstract}

The recent discovery of the third interstellar object (3I/ATLAS) expands the known census from two to three and significantly improves statistical inferences regarding the underlying galactic population. In this paper, we argue that cometary activity likely significantly contributes to 3I/ATLAS's brightness, since the nuclear size inferred when assuming an asteroidal reflectance implies an untenable interstellar object mass per star. 3I/ATLAS exhibits a high excess velocity of $v_\infty=58$~km/s relative to the Sun, which  implies that 3I/ATLAS is relatively old in comparison to previous interstellar objects. Here, we calculate the posterior distribution of ages implied by the kinematics of the interstellar objects and find that 3I/ATLAS is likely $\sim3-11$~Gyr old, assuming that the interstellar object and stellar age-velocity dispersion relations are equivalent. We also calculate the distribution of host star metallicities and find that 3I/ATLAS has a 12 \% chance of originating from a star with $\text{[Fe/H]}\leq-0.4$. These results show that interstellar object formation is likely efficient at low metallicities and early in the history of the Galaxy. Finally, we estimate the interstellar object formation rate throughout Galactic history implied by these three objects. As future interstellar objects are discovered, the framework presented here can be applied to further refine this calculation. Comparison between the interstellar object and stellar formation histories will provide unique insights into the history of stellar system formation in the Galaxy.

\end{abstract}


\section{Introduction} \label{sec:intro}

The third interstellar object, 3I/ATLAS, was recently discovered traversing through the solar system \citep{Denneau2025, Seligman2025}. This discovery follows the discoveries of 1I/`Oumuamua in 2017 \citep{MPEC2017, Williams2017} and 2I/Borisov in 2019 \citep{Borisov2019}. These objects {most likely} formed in protoplanetary disks {alongside exoplanets} and were subsequently ejected via planet scattering or post--main sequence mass loss \citep{PortegiesZwart2018, Do2018, Levine2023}. 

Although the system that produced a given interstellar object is impossible to determine \citep{Hallatt2020}, these objects provide a unique opportunity to directly observe material formed in extrasolar  systems. Given the difficulty of directly identifying an origin, statistical analyses are necessary to understand the population of interstellar objects. By going from two to three objects, the discovery of 3I/ATLAS expands this sample by \qty{50}{\percent} {and} improves the robustness of these sample-limited statistics. In an ensemble, the kinematics and chemodynamics of Galactic interstellar object streams will trace the history of Galactic star formation \citep{PortegiesZwart2021, Lintott2022, Hopkins2023, Forbes2025, Hopkins2025a}.

This discovery is especially relevant given the mysterious and divergent properties of the first two interstellar objects. In particular, 1I/`Oumuamua exhibited significant nongravitational acceleration \citep{Micheli2018} but had no visible activity \citep{Meech2017, Jewitt2017, Trilling2018}. 1I/`Oumuamua {also} exhibited significant brightness variations with a period of \qty{4.3}{hr} \citep{Belton2018}, indicating a rotating elongated body. Absolute magnitude variations further indicated that 1I/`Oumuamua was experiencing complex, nonprincipal axis rotation \citep{Drahus2017, Meech2017, Fraser2018, Taylor2023a} and had an extreme oblate shape \citep{Mashchenko2019}. 

The complex properties of 1I/`Oumuamua led to a wide variety of hypothesized origins. While some authors suggested that a low-density fractal aggregate \citep{Moro-Martin2019, Flekkoy2019} or an ultra-thin geometry \citep{Bialy2018} would allow radiation pressure to explain the nongravitational acceleration, others demonstrated that low dust production during outgassing could also be responsible (e.g., \citealt{Sekanina2019}).

While H$_2$O ice is the most common solar system comet volatile (e.g., \citealt{Ahearn2012, Cochran2015}) and is consistent with observations \citep{Trilling2018, Park2018}, energy constraints implied that H$_2$O could not be responsible for the nongravitational acceleration \citep{Sekanina2019}. Constraints on CO and CO$_2$ outgassing from the Spitzer Space Telescope \citep{Trilling2018} {further implied} that some other, more exotic volatile must be responsible for the nongravitational acceleration. Therefore, N$_2$ \citep{Jackson2021, Desch2021} or H$_2$ \citep{Seligman2020, Bergner2023} outgassing have also been proposed as solutions, although the provenance of such compositions are debated (see, e.g., \citealt{Seligman2020, Hsieh2021, Levine2021, Desch2022, Bergner2023}). {For example,} it has been suggested that 1I/`Oumuamua formed in giant molecular cloud instead {of a protoplanetary disk} \citep{Seligman2020, Hsieh2021, Levine2021}. 

Meanwhile, the properties of 2I/Borisov were more typical for a comet. The object {exhibited} distinct cometary activity and a volatile composition broadly similar to that of some solar system comets \citep{Xing2020, Lin2020, Bannister2020}{, although the} composition was enriched in CO relative to H$_2$O \citep{Bodewits2020, Cordiner2020, Kareta2020}. 2I/Borisov also exhibited a (predicted, \citealt{Jewitt2019a, Jewitt2019}) spin-up due to its outgassing, which lead to an outburst and disintegration event post-perihelion \citep{Jewitt2020}. 

While observations are still ongoing, 3I/ATLAS exhibits an active coma similar to that of 2I/Borisov \citep{Alarcon2025, Jewitt2025, Seligman2025, Opitom2025}, {including during precovery observations \citep{Feinstein2025}}{with tentative evidence of H$_2$O ice \citep{Yang2025}.} {T}he spectral slope of 3I/ATLAS is similar to that of 2I/Borisov and that of D-type asteroids and is consistent with that of 1I/`Oumuamua \citep{DeMeo2009, Seligman2025, Puzia2025, Santana-Ros2025}. {It may be feasible to send spacecraft from Mars to fly by 3I/ATLAS and study its physical properties \textit{in situ}, which would provide stricter constraints on the composition of this object \citep{Yaginuma2025}. } 

In addition, the excess velocities of 1I/`Oumuamua ($v_\infty=\qty{26}{km/s}$) and 2I/Borisov ($v_\infty=\qty{32}{km/s}$) indicated distinct ages of $\sim$\qty{100}{Myr} and $\sim$\qty{1}{Gyr}, respectively \citep{Mamajek2017, Gaidos2017, Feng2018, Hallatt2020, Hsieh2021}. 3I/ATLAS is a significant outlier in this respect, with an excess velocity of $v_\infty=\qty{58}{km/s}$. A similar scaling would indicate that 3I/ATLAS has an age of $\sim$\qty{10}{Gyr}, although this is based on a rough comparison between excess velocity and stellar velocity dispersion and is inherently approximate. Our current understanding of 1I/`Oumuamua and 2I/Borisov has been reviewed in \citet{Moro-Martin2022a, Fitzsimmons2023, Seligman2023a}, and \citet{Jewitt2023}.

In this work, we provide interstellar object population calculations  incorporating 3I/ATLAS. In Sec. \ref{sec:numden}, we argue that 3I/ATLAS must have a smaller size than currently measured, somewhat comparable to that of 2I/Borisov. Next, we consider the Galactic velocity radiants of the three interstellar objects and calculate the probability density functions of their ages implied by these velocities in Sec. \ref{sec:ages}. In Sec. \ref{sec:metals}, we calculate the distribution of {parent} stellar metallicities that these ages imply. We then approximate the interstellar object formation rate in Sec. \ref{sec:ISOFR} and finally conclude and discuss our results in Sec. \ref{sec:discussion}.

\section{Estimated Number Density and Size}\label{sec:numden}

{Here, we estimate the number density of objects with the size of 3I/ATLAS and determine that it must be smaller than the calculated upper limit. Assuming a reasonable interstellar object production rate, we infer that this object should have a size $D\simeq\qty{2}{km}$ and a significant coma. A similar calculation was performed by \citet{Loeb2025}.}

{First,} we calculate an approximate number density by assuming that the detection rate for an object like 3I/ATLAS is $\Gamma=\qty{0.1}{yr^{-1}}$. This value reflects the time that the ATLAS survey has been online, especially when combined with the Pan-STARRS1 survey \citep{Chambers2016} and the Catalina Sky Survey, which are also sensitive to objects of this magnitude. Now, the ATLAS survey {observes nearly the entire night sky down to an apparent magnitude of 19, where the survey has a \qty{75}{\percent} completeness \citep{Tonry2018}}. Since 3I/ATLAS has an absolute magnitude of $H\simeq12$, an object with $H\leq12$ is detectable by ATLAS at a distance of approximately $d=\qty{5}{au}$, {which} corresponds to a detection cross-section of $\sigma=\pi d^2$. {We have assumed that the observed object is equidistant from both the Earth and the Sun at a distance $d$ and that the brightness is reduced by an average of \num{0.3}, which roughly accounts for the phase angle of the small body.} Since 3I/ATLAS is moving at $v_\infty=\qty{58}{km/s}$, the total number density is $n=\Gamma/(\sigma v_\infty)\simeq\qty{e-4}{au^{-3}}$. This value is comparable to the previous estimate of \qty{3e-4}{au^{-3}} \citep{Seligman2025} if slightly different assumptions are made about the ATLAS detection area. 

However, this value implies an exceptionally high mass of interstellar objects produced per star. If the stars in the Galaxy {have a number density of $n\sim\qty{0.1}{pc^3}$ \citep{Bovy2017}}, then each star should eject \num{e13} interstellar objects the size of 3I/ATLAS or larger. If 3I/ATLAS were nearly $r_N\simeq\qty{10}{km}$ across {(assuming asteroidal reflectance and an albedo of $p=0.05$, see \citealt{Seligman2025})}, at the typical cometary density of $\rho=\qty{1}{g/cm^3}$ each star must produce approximately \qty{e-2}{M_\odot} of interstellar objects larger than 3I/ATLAS. There must be some additional mass stored in smaller objects, which will only add to this (already significant) value. 

It is presumably infeasible for formation processes to produce this {abundance} of interstellar objects. Since the typical star is \qty{0.5}{M_\odot}, this would imply that there is $\sim50$ times the mass in stars than in interstellar objects. {However, interstellar objects are made of metals, while stars are mostly hydrogen. If we assume that the Galactic metallicity is approximately solar ($Z=\num{0.015}$, \citealt{Asplund2021})}, interstellar object {formation requires nearly twice as much mass as star formation}. 

The true size of 3I/ATLAS must {therefore} be somewhat less than the estimated \qty{10}{km} upper limit, {as expected for an active object like 3I/ATLAS.} In order for 3I/ATLAS to imply a {more reasonable} mass of $M_{\rm ISO}\simeq\qty{100}{M_\oplus}$ in interstellar objects per star, 3I/ATLAS must have a radius $r_N$ of (see also \citealt{Loeb2025})
\begin{equation}
\begin{split}
    r_N=\qty{2.3}{km}&\left(\frac{M_{\rm ISO}}{\qty{100}{M_\oplus}}\right)^{1/3}\left(\frac{\rho}{\qty{1}{g/cm^3}}\right)^{-1/3}\,.
\end{split}
\end{equation}
Given the activity already detected on 3I/ATLAS \citep{Alarcon2025, Jewitt2025, Seligman2025, Opitom2025, delaFuenteMarcos2025, Chandler2025}, this is a somewhat reasonable value, since the coma can account for a significant fraction of its brightness. If 3I/ATLAS's nuclear size is closer to \qty{2}{km}, then the body itself will have an absolute magnitude of $H=15.5$. Given that the measured absolute magnitude is $H=12.4$, the coma {would then} account for $\sim$\qty{94}{\percent} of 3I/ATLAS's flux. {Later Hubble observations have demonstrated that the nuclear radius of 3I/ATLAS must be between \qty{0.16}{km} and \qty{2.8}{km} \citep{Jewitt2025a}. This upper limit is nicely in line with our predicted size. }

\section{Kinematic Age}\label{sec:ages}

The excess velocity of 3I/ATLAS is nearly $v_\infty\simeq\qty{58}{km/s}$ relative to the Sun, significantly higher than 1I/`Oumuamua's velocity of \qty{26}{km/s} or 2I/Borisov's \qty{32}{km/s}. This high velocity implies a similarly high age for 3I/ATLAS relative to the other two interstellar objects. 

For a given measured velocity, we can calculate the probability density function (PDF) of the expected ages {given an age-velocity relation}. We first assume that interstellar objects have a velocity distribution close to that of their parent star. {This circumstance} is expected, since the stellar velocity distribution is modified over time by the effects of the Galactic potential, which will have nearly identical effects on the interstellar objects \citep{Seligman2018}. With this assumption, the age PDF of an interstellar object with a Galactic velocity $U$, $V$, $W$ is equivalent to that of a star with the same velocity. 

There exists previous work (e.g., \citealt{Almeida-Fernandes2018a}) calculating the age of stars based on their Galactic velocities, which we will briefly summarize here. Consider a star with a Galactic velocity $U$, $V$, $W$ relative to the Sun. This velocity can be converted to the independent velocity ellipsoid values $v_1$ and $v_2$, which ignore the correlation between $U$ and $V$. Consider a single velocity value $v_i$. {We assume that} stars of a given age $\tau$ will have a Gaussian distribution of $v_i$ velocities with a mean of zero relative to the LSR and a velocity dispersion $\sigma_i(\tau)$. {While the stellar velocity distribution is dominated by moving groups \citep{Dehnen1998, Ramos2018}, it can be approximated by a Gaussian.} {With this assumption}, the probability of obtaining a velocity $v_i$ from a star of age $\tau$ is given by 
\begin{equation}
    p(v_i|\tau)=\frac{1}{\sqrt{2\pi}\sigma_i(\tau)}\exp\left(-\frac{v_i^2}{2\sigma_i^2(\tau)}\right)\,.
\end{equation}

Using Bayes' theorem, the {posterior} age probability distribution $p(\tau|v_1,v_2,v_3)$ is given by 
\begin{equation}\label{eq:ptau}
    p(\tau|v_1,v_2,v_3)=p(\tau)\prod_i p(v_i|\tau)\,,
\end{equation}
where $p(\tau)$ is the {prior} probability of finding a star with age $\tau$. {The posterior distribution} is normalized such that $\int p(\tau|v_1,v_2,v_3)\de\tau=1$. {Since the star formation history of the Milky Way is approximately constant  \citep{Snaith2015, Xiang2022}, we can safely assume a flat prior in age.} Previous work \citep{Almeida-Fernandes2018a} used the Geneva-Copenhagen Survey \citep{Nordstrom2004, Holmberg2009}, an all-sky survey of F and G dwarfs in the solar neighborhood, to calibrate the relationship between the stellar kinematics and the ages. {They found that for} a given $U$, $V$, and $W$, 
\begin{subequations}
\begin{align}
    v_1=&(U+U_\odot)\cos l_v(\tau)+(V+V_\odot'(\tau))\sin l_v(\tau)\,;\\
    v_2=&(V+V_\odot'(\tau))\cos l_v(\tau) - (U+U_\odot)\sin l_v(\tau)\,;\\
    v_3=&W+W_\odot\,.
\end{align}
\end{subequations}
Here, $U_\odot=\qty{9.8}{km/s}$, $W_\odot=\qty{7.2}{km/s}$, 
\begin{equation}
    l_v(\tau)=0.41\exp(-0.37(\tau/\unit{Gyr}))\,,
\end{equation}
and 
\begin{equation}
    V_\odot'(\tau)=0.17\left(\frac{\tau}{\unit{Gyr}}\right)^2+0.63\left(\frac{\tau}{\unit{Gyr}}\right)+12.5\,.
\end{equation}
Finally, 
\begin{subequations}\label{eq:sigmas}
\begin{align}
    \sigma_1(\tau)=&22.0\left(\frac{\tau}{\unit{Gyr}}\right)^{0.33}\,,\\
    \sigma_2(\tau)=&11.9\left(\frac{\tau}{\unit{Gyr}}\right)^{0.42}\,,\text{ and}\\
    \sigma_3(\tau)=&9.1\left(\frac{\tau}{\unit{Gyr}}\right)^{0.48}\,.
\end{align}
\end{subequations}
{The values in Eq. \eqref{eq:sigmas} are taken from Table 3 in \citet{Almeida-Fernandes2018}.}

\begin{table*}[t]
    \centering
    \caption{\textbf{Galactic Interstellar Object Velocities.} The Galactic velocities of the interstellar objects. }
    \begin{tabular}{l|llll}
         Object & $U$ [\unit{km/s}] & $V$ [\unit{km/s}] & $W$ [\unit{km/s}] & Citation\\\hline
         1I/`Oumuamua & -11.457 & -22.395 & -7.746 & \citealt{Mamajek2017} \\
         2I/Borisov & \phantom{-}22.000 & -23.601 & \phantom{-}1.057 & \citealt{deLeon2020} \\
         3I/ATLAS & -51.14 & -19.33 & \phantom{-}18.86 & \citealt{Seligman2025}
    \end{tabular}
    \label{tab:ISOdata}
\end{table*}

\begin{figure}
    \centering
    \includegraphics{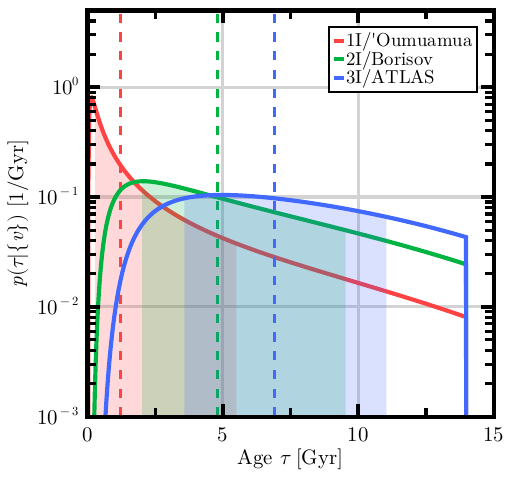}
    \caption{\textbf{Interstellar Object Age.} The {posterior} probability distribution function $p(\tau|\{v\}_i)$ for the three known interstellar objects{, where $\{v\}$ represents the set of measured interstellar object velocities}. We show the age distribution as well as the median (dashed line) and the \qty{68}{\percent} confidence region (shaded).}
    \label{fig:iso_age}
\end{figure}

The velocities of the three interstellar objects in Galactic coordinates are given in Table \ref{tab:ISOdata}. We {then calculate the posterior probability distribution of the ages of the three interstellar objects, which we show in Fig. \ref{fig:iso_age}.} We also show the median (vertical dashed lines) and the \qty{68}{\percent} confidence regions (shaded) {of the age values}. {This confidence interval is calculated to be the region between the 16th and 84th percentile of the age PDF.} We find that the {median} age of 1I/`Oumuamua is $\tau_1=1.2^{+4.3}_{-1.0}$ Gyr, the {median} age of 2I/Borisov is $\tau_2=4.8^{+4.7}_{-2.8}$ Gyr, and the {median} age of 3I/ATLAS is $\tau_3=6.9^{+4.1}_{-3.3}$ Gyr. 

As expected, the higher the velocity of the interstellar object, the higher its age. However, the distribution of ages generally overlaps far more than expected. Even 1I/`Oumuamua, the youngest interstellar object, has a median age significantly higher than the \qtyrange{30}{100}{Myr} generally discussed in the literature \citep{Feng2018, Hsieh2021}. This age is broadly consistent with a previous kinematical age estimate of \qtyrange{0.01}{1.87}{Gyr} \citep{Almeida-Fernandes2018}{, although those authors appear to have mistakenly used $p(\tau)$ in place of $p(\log\tau)$ when calculating the confidence intervals and thus have underestimated the age.}

Regardless, the expected age of 3I/ATLAS is approximately \qtyrange{3}{11}{Gyr}. This is a significant range, but indicates that this object is old and is likely significantly older than 1I/`Oumuamua or 2I/Borisov. This age is also consistent with the values predicted by the \=Otautahi-Oxford population model \citep{Hopkins2025}. 

\begin{figure}
    \centering
    \includegraphics{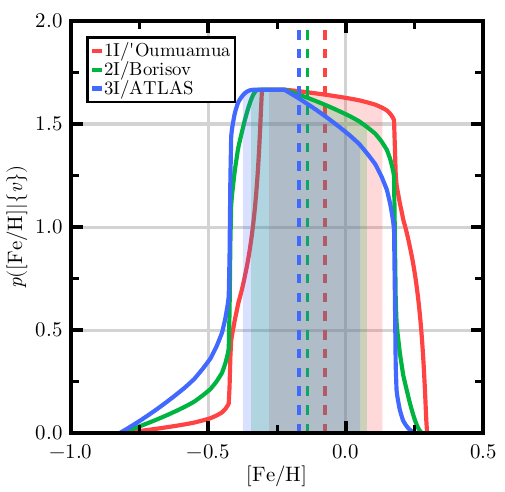}
    \caption{\textbf{Interstellar Object Metallicity.} The probability density function $p(\text{[Fe/H]})$ for the interstellar objects as a function of the metallicity. The median is shown as a dashed line and the \qty{68}{\percent} confidence region is shaded.}
    \label{fig:iso_feh}
\end{figure}

\section{Parent Stellar Metallicities}\label{sec:metals}

Given that the Galactic metallicity increases with time, we can convert these age probability distributions into metallicity distributions. For a given age, we assume that the metallicity distribution has a uniform scatter of $\pm\qty{0.3}{dex}$ and a median value $\overline{\feh}(\tau)$ given by Fig. 2 in \citet{Marsakov2011}. Assuming that the metallicity follows a uniform distribution provides
\begin{equation}\label{eq:fehprior}
    p(\feh|\tau)=\begin{cases}
        5/3 & |\feh-\overline{\feh}(\tau)|\leq\qty{0.3}{dex};\\
        0 & \text{otherwise}\,.
    \end{cases}
\end{equation} 
{The posterior distribution is therefore given by} 
\begin{equation}
    p(\feh|\{v\})=\int p(\feh|\tau)p(\tau|\{v\})\de\tau\,.
\end{equation}
With $p(\tau|\{v\})$ given by Eq. \eqref{eq:ptau} and $p(\feh|\tau)$ given by Eq. \eqref{eq:fehprior}, we calculate $p(\feh|\{v\})$, {which is shown} in Fig. \ref{fig:iso_feh}. The median parent metallicity of 1I/`Oumuamua is $\text{[Fe/H]}=-0.08\pm0.21$, 2I/Borisov has $\text{[Fe/H]}=-0.15^{+0.22}_{-0.21}$, and 3I/ATLAS has $\text{[Fe/H]}=-0.18^{+0.23}_{-0.21}$. {In this model, there is a \qty{12}{\percent} chance that 3I/ATLAS came from a star with a metallicity $\feh\leq-0.4$ and a \qty{5}{\percent} chance it came from a star with $\feh\leq-0.5$. These values are somewhat counter to previous predictions, which have low probabilities of finding interstellar objects from low-metallicity stars \citep{Hopkins2025a}. However, these calculations use theoretical predictions of the interstellar object formation rate as a function of stellar type and metallicity as a prior, while we calculate the metallicity distribution from observations alone. }

While the three interstellar objects do not exhibit significant variation in their parent stellar population metallicity distribution, the older objects likely originated from lower-metallicity systems. Specifically, it is not unreasonable for 3I/ATLAS to be formed around a star with a metallicity as low as \qty{-0.4}{dex}, which would represent a system with fewer metals than the Sun by a factor of $0.4$. This result implies that interstellar object formation processes are operating at metallicities somewhat lower than the Sun. The detection of planets around stars of similar metallicity ($\text{[Fe/H]}=-0.68$, \citealt{Cochran2007}) is consistent with this result.  

\begin{figure}
    \centering
    \includegraphics{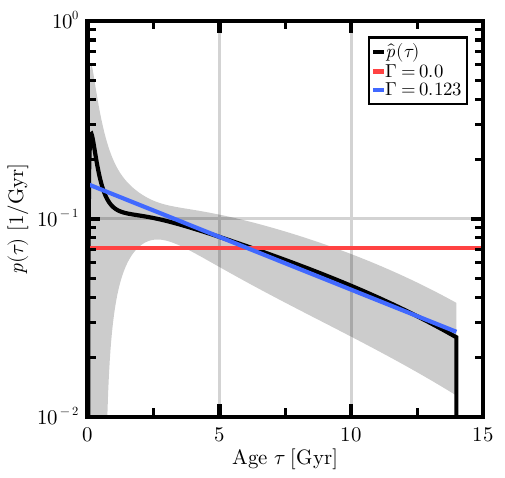}
    \caption{\textbf{{Inferred }Interstellar Object {Prior}.} An approximate interstellar object {local age distribution}, assuming that the weight for each interstellar object is equal. {The $1\sigma$ uncertainty is shown as a shadowed region. We also show the expected local age distributions assuming that ISOs are not destroyed ($\Gamma=0$) and are uniformly destroyed at a rate $\Gamma=\qty{0.123}{Gyr^{-1}}$.}}
    \label{fig:iso_prod}
\end{figure}

\section{Interstellar Object Formation Rate}\label{sec:ISOFR}

The fact that these interstellar objects cover a range of possible age values implies that interstellar objects must be formed over a similarly wide range of ages. Indeed, we can use the expected age PDFs to estimate the production rate of interstellar objects over time. 

{The age distribution of interstellar objects near the Sun will be equivalent to the prior distribution of measured ages.} Given $N$ observations, each of which has a PDF $p_i(\tau|\{v\})$, {we marginalize over these PDFs to find the estimated prior PDF}
\begin{equation}\label{eq:ptauest}
    \hat{p}(\tau)=\sum_iw_ip_i(\tau|\{v\})\,,
\end{equation}
where $w_i$ are the weights of the observations and $\sum w_i=1$. {Eq. \eqref{eq:ptauest} is an unbiased estimator for $p(\tau)$, marginalized over the velocities.} The weight of each observation is {equivalent to} the probability of such an observation occurring.  { Since the age estimations in this case depend only on the measured velocity, the weights will depend on each telescope's completeness at the observed velocity. While this is difficult to estimate, it is not unreasonable to suggest that the detectability of an interstellar object is roughly independent of its velocity up to a certain point, although it has been suggested that interstellar objects with slow sky-plane velocities  have been preferentially detected to date (see Figure 2 in \citealt{Marceta2023}). If this assumption is accurate to within an order of magnitude, then for $N$ observations the weights will be $w_i=1/N$.} {Using these weights}, we {estimate} the total interstellar object {prior} $\hat{p}(\tau)$, which is shown in Fig. \ref{fig:iso_prod}.\footnote{{Note that for this calculation to converge to the actual prior as $N\rightarrow\infty$, we must use flat, uninformative priors to calculate the age distributions.}} The uncertainty in $\hat{p}(\tau)$ is given by {the square root of an unbiased estimator of the variance in $p(\tau)$}
\begin{equation}
    \sigma(\tau)=\sqrt{\frac{\sum_i(p_i(\tau)-\hat{p}(\tau))^2}{N(N-1)}}\,.
\end{equation}
{The $1\sigma$ interval is also shown in Fig. \ref{fig:iso_prod}.}

{The distribution of interstellar object ages is not necessarily proportional to the production rate of interstellar objects. Let us assume that interstellar objects are produced at a rate $S(t)$ and have an age-independent probability $\Gamma(t)$ of being destroyed at a given time $t$. For a population with number $N(t)$, the loss rate of objects at a time $t$ is given by $\dot{N}(t)=-\Gamma(t)N(t)$. Rearranging, we find that the total probability of an object with an age $\tau$ being destroyed is}
\begin{equation}
    f_{\rm des}(\tau)=\exp\Big(-\!\!\int\displaylimits_0^\tau\!\Gamma(t)\,\de t\Big)\,.
\end{equation}
{Now, the population with an age $\tau$ is proportional to the production rate $S(t)$. Therefore, the age distribution of objects at the current time is}
\begin{equation}
    p(\tau)\propto S(\tau)\exp\Big(-\!\!\int\displaylimits_0^\tau\!\Gamma(t)\,\de t\Big)\,,
\end{equation}
{where $p(\tau)$ is normalized such that $\int p(\tau)\de\tau=1$. }

{We consider two possible cases for the destruction and production rate. First, we assume that the destruction rate $\Gamma(t)=0$ and the production rate $S(t)=\lambda$, a constant. This example is shown in Fig. \ref{fig:iso_prod}. In this case, $p(\tau)\propto S(\tau)$, so the calculated prior is proportional to the interstellar object production rate. To determine if this distribution is consistent with the calculated production rate $\hat{p}(\tau)$, we generate \num{e4} realizations of the $\hat{p}(\tau)$ distribution from its calculated uncertainty distribution. For each realization, we compare the distributions using a two-sample Kolmogorov-Smirnov (K-S) test \citep{Massey1951}. The maximum $p$-value is $p\sim\num{e-16}$, so this distribution is entirely inconsistent with the $\Gamma=0$ curve.}

{Second, we assume that $S(t)=\lambda$ and $\Gamma(t)=\Gamma$, so that both are constant. With these assumptions, $p(\tau)=\Gamma\exp(-\Gamma\tau)/(1-\exp(-t_H\Gamma))$, where $t_H=\qty{13.8}{Gyr}$ is the age of the universe. In this case, we can fit this distribution to the median empirical $\hat{p}(\tau)$ distribution and find that $\Gamma=\qty{0.123}{Gyr^{-1}}$ minimizes the integrated square error. When we compare these distributions with a suite of K-S tests, we find that the mean $p$-value is $p=0.06$, so this distribution remains consistent with a flat production and destruction rate.  }

{While the distribution is consistent with a uniform destruction rate, the destruction rate of interstellar objects is generally assumed to be negligible in comparison to production \citep{Hopkins2023, Hopkins2025a}, {and there is no known mechanism to destroy interstellar objects so efficiently across the Galaxy. Assuming that interstellar objects are not destroyed } ($\Gamma=0$), the detection of the three currently-known interstellar objects is entirely inconsistent with a flat interstellar object production rate, and a gradually increasing production rate is necessary {to explain the data}.} Given these assumptions, Fig. \ref{fig:iso_prod} indicates that the interstellar object production rates {rose steadily} for the first \qty{10}{Gyr} of Galactic history {and sharply} increased in the last \qty{100}{Myr} {(although there are significant uncertainties at recent times). As more interstellar objects are discovered, this calculation provides a framework for determining the production history of interstellar objects in the Galaxy.} 

{We also must note that we expect a detection bias against younger interstellar objects. The combination of a young age and a low velocity means that these objects will have a limited range from their parent star and are unlikely to be detected. For example,} interstellar objects with ages of $\tau\sim\qty{10}{Myr}$ have typical velocities of only \qty{5}{km/s} and have a maximum range of \qty{50}{pc}. {The sample of interstellar objects with this age would therefore be volume-limited}, since many interstellar objects with ages $\tau\leq\qty{10}{Myr}$ could not have made it to the solar system to be detected. 

\section{Discussion}\label{sec:discussion}

In this paper, we have built off previous work on the ages of 1I/`Oumuamua and 2I/Borisov (e.g., \citealt{Almeida-Fernandes2018, Hallatt2020}) to calculate the kinematic age of the new {interstellar} object 3I/ATLAS. Our results indicate that 3I/ATLAS is the oldest detected interstellar object, approximately \qtyrange{3}{11}{Gyr} old. 3I/ATLAS therefore provides a sample of planetary system formation in the early Galaxy, confirming that such systems exist and allowing us to determine how such systems may differ from the younger exoplanet systems currently studied. Given its age, this object also {may have} originated in the thick disk of the Galaxy, providing a sample of system formation in this region. 

We have also calculated the metallicity distribution for the stars that may have formed each interstellar object. As expected, the older objects generally favor lower-metallicity parents, although all metallicity distributions are similar and relatively flat. Due to the large inherent scatter in the age-metallicity relation, the interstellar object metallicities are very broad and cannot be narrowed down for such a small sample.  It may be possible to determine if interstellar objects preferentially formed around stars of a certain metallicity when future interstellar objects are discovered. However, this will require a larger sample than is currently available. 

In addition, the planetesimal mass budget at low metallicities is relatively limited \citep{Andama2024}. If the size of 3I/ATLAS is $r_N=\qty{2.3}{km}$, then the interstellar object mass budget for systems with a metallicity of $\text{[Fe/H]}\simeq\numrange{-0.4}{0.03}$ is 
\begin{equation}
    M_{\rm ISO}=\qty{100}{M_\oplus}\left(\frac{r_N}{\qty{2.3}{km}}\right)^3\,.
\end{equation}
This predicted mass budget will vary based on the size of 3I/ATLAS. Once the size of this object is established, it will be possible to estimate the corresponding mass budget.

Regardless, we have shown that 3I/ATLAS is old and likely formed around a low-metallicity star. The detection of this object therefore indicates that interstellar object formation is efficient at low metallicities and early in the history of the Galaxy. While planets have been found around low-metallicity stars (e.g., \citealt{Cochran2007}), the detection of 3I/ATLAS provides an independent confirmation of this result. 

Lastly, we calculated the {local} interstellar object {age prior} implied by these detections{, which is closely related to the interstellar object formation rate}. Our calculation suggests that while interstellar object production occurred early in the lifetime of the Galaxy, the interstellar object formation rate increased for the first {\qty{10}{Gyr}}. The interstellar object formation rate then increased to a peak in the last $\sim$\qty{300}{Myr}{, although the estimated prior in this region has large uncertainties. We have also shown that this prior is inconsistent with a flat interstellar object formation rate, assuming that negligible interstellar objects are destroyed in the past. However, the implied prior is consistent with a uniform destruction fraction of $\Gamma=\qty{0.123}{Gyr^{-1}}$, so that \qty{12}{\percent} of interstellar objects are destroyed in a given \unit{Gyr}. {There is no known mechanism that can produce such a significant destruction rate of interstellar objects. Stellar encounter rates are as low as \qty{e-7}{Gyr^{-1}} and only a small fraction of those encounters will lead to destruction \citep{Forbes2019, Forbes2025}. In addition, while there will be $\sim\num{e7}$ supernovae in the Galaxy during one \unit{Gyr}, they will not be able to deliver sufficient energy to destroy macroscopic interstellar objects to the requisite \qty{12}{\percent} of the Galaxy.} If interstellar objects are not destroyed, then this result is instead consistent with an exponentially increasing interstellar object production rate through Galactic history. }

Finally, we describe caveats to these results. Our calculations are preliminary in nature and heavily rely on  approximation. While many of these approximations are appropriate and will hold to an order-of-magnitude level, our results will change as more interstellar objects are detected and as the physical properties of 3I/ATLAS are better constrained with forthcoming observations. {Particular care must be taken when interpreting our inferred interstellar object prior/formation rate, since small-number statistics introduces relatively large uncertainties. Finally, the uncertainty on the destruction rate of interstellar objects per unit time makes it difficult to accurately estimate the interstellar object formation history of the Galaxy.}

Fortunately, we expect the recently-commissioned Vera C. Rubin Observatory and its Legacy Survey of Space and Time (LSST) to detect many more interstellar objects \citep{Hoover2022,Marceta2023,Dorsey2025}. As the population sample grows, statistical calculations will become more robust and converge. Forthcoming observations of 3I/ATLAS will provide insight into an object that may have formed in a stellar system nearly as old as the Galaxy itself. {As more interstellar objects are detected, the uncertainties in this estimation will be reduced and the interstellar object formation rate will become clearer. Once formation rate estimates are robust, comparison between the interstellar object formation rate and the Galactic star formation rate will provide estimates of the number of interstellar objects produced per star throughout Galactic history. Such an estimate will provide meaningful insight into the history of planetesimal and planet formation early in the Galactic history. }

\section*{Acknowledgments}

{We thank the anonymous reviewer for their useful comments.} We thank Gregory Laughlin and {Fred Adams} for insightful conversations. A.G.T. acknowledges support from the Fannie and John Hertz Foundation and the University of Michigan's Rackham Merit Fellowship Program. D.Z.S. is supported by an NSF Astronomy and Astrophysics Postdoctoral Fellowship under award AST-2303553. This research award is partially funded by a generous gift of Charles Simonyi to the NSF Division of Astronomical Sciences. The award is made in recognition of significant contributions to Rubin Observatory’s Legacy Survey of Space and Time. 
    
\bibliography{main}{}

\begin{thebibliography}{}
\expandafter\ifx\csname natexlab\endcsname\relax\def\natexlab#1{#1}\fi
\providecommand{\url}[1]{\href{#1}{#1}}
\providecommand{\dodoi}[1]{doi:~\href{http://doi.org/#1}{\nolinkurl{#1}}}
\providecommand{\doeprint}[1]{\href{http://ascl.net/#1}{\nolinkurl{http://ascl.net/#1}}}
\providecommand{\doarXiv}[1]{\href{https://arxiv.org/abs/#1}{\nolinkurl{https://arxiv.org/abs/#1}}}

\bibitem[{M.~F. A'Hearn {et~al.}(2012)A'Hearn, Feaga, Keller, Kawakita, Hampton, Kissel, Klaasen, McFadden, Meech, Schultz, Sunshine, Thomas, Veverka, Yeomans, Besse, Bodewits, Farnham, Groussin, Kelley, Lisse, Merlin, Protopapa, \& Wellnitz}]{Ahearn2012}
A'Hearn, M.~F., Feaga, L.~M., Keller, H.~U., {et~al.} 2012, \bibinfo{title}{{COMETARY} {VOLATILES} {AND} {THE} {ORIGIN} {OF} {COMETS},} The Astrophysical Journal, 758, 29, \dodoi{10.1088/0004-637X/758/1/29}

\bibitem[{M.~R. Alarcon {et~al.}(2025)Alarcon, Serra-Ricart, Licandro, Guerra~Arrencibia, Ruiz~Cejudo, \& Trujillo}]{Alarcon2025}
Alarcon, M.~R., Serra-Ricart, M., Licandro, J., {et~al.} 2025, \bibinfo{title}{Deep g'-band {Imaging} of {Interstellar} {Comet} {3I}/{ATLAS} from the {Two}-meter {Twin} {Telescope} ({TTT}),} \{The Astronomer's Telegram\} No. 17263

\bibitem[{F. Almeida-Fernandes \& H.~J. Rocha-Pinto(2018{\natexlab{a}})Almeida-Fernandes \& Rocha-Pinto}]{Almeida-Fernandes2018a}
Almeida-Fernandes, F., \& Rocha-Pinto, H.~J. 2018{\natexlab{a}}, \bibinfo{title}{A method to estimate stellar ages from kinematical data,} Monthly Notices of the Royal Astronomical Society, 476, 184, \dodoi{10.1093/mnras/sty119}

\bibitem[{F. Almeida-Fernandes \& H.~J. Rocha-Pinto(2018{\natexlab{b}})Almeida-Fernandes \& Rocha-Pinto}]{Almeida-Fernandes2018}
Almeida-Fernandes, F., \& Rocha-Pinto, H.~J. 2018{\natexlab{b}}, \bibinfo{title}{A kinematical age for the interstellar object {1I}/‘{Oumuamua},} Monthly Notices of the Royal Astronomical Society, 480, 4903, \dodoi{10.1093/mnras/sty2202}

\bibitem[{G. Andama {et~al.}(2024)Andama, Mah, \& Bitsch}]{Andama2024}
Andama, G., Mah, J., \& Bitsch, B. 2024, \bibinfo{title}{Which stars can form planets: {Planetesimal} formation at low metallicities,} Astronomy \& Astrophysics, 683, A118, \dodoi{10.1051/0004-6361/202348899}

\bibitem[{M. Asplund {et~al.}(2021)Asplund, Amarsi, \& Grevesse}]{Asplund2021}
Asplund, M., Amarsi, A.~M., \& Grevesse, N. 2021, \bibinfo{title}{The chemical make-up of the {Sun}: {A} 2020 vision,} Astronomy \& Astrophysics, 653, A141, \dodoi{10.1051/0004-6361/202140445}

\bibitem[{M.~T. Bannister {et~al.}(2020)Bannister, Opitom, Fitzsimmons, Moulane, Jehin, Seligman, Rousselot, Knight, Marsset, Schwamb, Guilbert-Lepoutre, Jorda, Vernazza, \& Benkhaldoun}]{Bannister2020}
Bannister, M.~T., Opitom, C., Fitzsimmons, A., {et~al.} 2020, \bibinfo{title}{Interstellar comet {2I}/borisov as seen by {MUSE}: {C}₂, {NH}₂ and red {CN} detections,} arXiv e-prints, arXiv:2001.11605

\bibitem[{M.~J.~S. Belton {et~al.}(2018)Belton, Hainaut, Meech, Mueller, Kleyna, Weaver, Buie, Drahus, Guzik, Wainscoat, Waniak, Handzlik, Kurowski, Xu, Sheppard, Micheli, Ebeling, \& Keane}]{Belton2018}
Belton, M. J.~S., Hainaut, O.~R., Meech, K.~J., {et~al.} 2018, \bibinfo{title}{The excited spin state of {1I}/2017 {U1} oumuamua,} ApJL, 856, L21, \dodoi{10.3847/2041-8213/aab370}

\bibitem[{J.~B. Bergner \& D.~Z. Seligman(2023)Bergner \& Seligman}]{Bergner2023}
Bergner, J.~B., \& Seligman, D.~Z. 2023, \bibinfo{title}{Acceleration of {1I}/‘{Oumuamua} from radiolytically produced {H2} in {H2O} ice,} Nature, 615, 610, \dodoi{10.1038/s41586-022-05687-w}

\bibitem[{S. Bialy \& A. Loeb(2018)Bialy \& Loeb}]{Bialy2018}
Bialy, S., \& Loeb, A. 2018, \bibinfo{title}{Could {Solar} {Radiation} {Pressure} {Explain} ‘{Oumuamua}’s {Peculiar} {Acceleration}?} The Astrophysical Journal Letters, 868, L1, \dodoi{10.3847/2041-8213/aaeda8}

\bibitem[{D. Bodewits {et~al.}(2020)Bodewits, Noonan, Feldman, Bannister, Farnocchia, Harris, Li, Mandt, Parker, \& Xing}]{Bodewits2020}
Bodewits, D., Noonan, J.~W., Feldman, P.~D., {et~al.} 2020, \bibinfo{title}{The carbon monoxide-rich interstellar comet {2I}/{Borisov},} Nature Astronomy, 4, 867, \dodoi{10.1038/s41550-020-1095-2}

\bibitem[{G. Borisov {et~al.}(2019)Borisov, Durig, Sato, Birtwhistle, Chen, Green, Bacci, Maestripieri, \& Nakano}]{Borisov2019}
Borisov, G., Durig, D.~T., Sato, H., {et~al.} 2019, \bibinfo{title}{Comet {C}/2019 {Q4} (borisov),} Central Bureau Electronic Telegrams, 4666, 1

\bibitem[{J. Bovy(2017)Bovy}]{Bovy2017}
Bovy, J. 2017, \bibinfo{title}{Stellar inventory of the solar neighbourhood using {Gaia} {DR1},} Monthly Notices of the Royal Astronomical Society, 470, 1360, \dodoi{10.1093/mnras/stx1277}

\bibitem[{K.~C. Chambers {et~al.}(2016)Chambers, Magnier, Metcalfe, Flewelling, Huber, Waters, Denneau, Draper, Farrow, Finkbeiner, Holmberg, Koppenhoefer, Price, Rest, Saglia, Schlafly, Smartt, Sweeney, Wainscoat, Burgett, Chastel, Grav, Heasley, Hodapp, Jedicke, Kaiser, Kudritzki, Luppino, Lupton, Monet, Morgan, Onaka, Shiao, Stubbs, Tonry, White, Bañados, Bell, Bender, Bernard, Boegner, Boffi, Botticella, Calamida, Casertano, Chen, Chen, Cole, Deacon, Frenk, Fitzsimmons, Gezari, Gibbs, Goessl, Goggia, Gourgue, Goldman, Grant, Grebel, Hambly, Hasinger, Heavens, Heckman, Henderson, Henning, Holman, Hopp, Ip, Isani, Jackson, Keyes, Koekemoer, Kotak, Le, Liska, Long, Lucey, Liu, Martin, Masci, McLean, Mindel, Misra, Morganson, Murphy, Obaika, Narayan, Nieto-Santisteban, Norberg, Peacock, Pier, Postman, Primak, Rae, Rai, Riess, Riffeser, Rix, Röser, Russel, Rutz, Schilbach, Schultz, Scolnic, Strolger, Szalay, Seitz, Small, Smith, Soderblom, Taylor, Thomson, Taylor, Thakar, Thiel, Thilker, Unger, Urata,
  Valenti, Wagner, Walder, Walter, Watters, Werner, Wood-Vasey, \& Wyse}]{Chambers2016}
Chambers, K.~C., Magnier, E.~A., Metcalfe, N., {et~al.} 2016, \bibinfo{title}{The pan-{STARRS1} surveys,} arXiv e-prints, arXiv:1612.05560, \dodoi{10.48550/arXiv.1612.05560}

\bibitem[{C.~O. Chandler {et~al.}(2025)Chandler, Bernardinelli, Jurić, Singh, Hsieh, Sullivan, Jones, Kurlander, Vavilov, Eggl, Holman, Spoto, Schwamb, Christensen, Beebe, Roodman, Lim, Jenness, Bosch, Smart, Bellm, MacBride, Rawls, Greenstreet, Slater, Heinze, Ivezić, Blum, Connolly, Daues, Makadia, Gower, Bryce~Kalmbach, Monet, Bannister, Dones, Dorsey, Fraser, Forbes, Fuentes, Holt, Inno, Jones, Knight, Lintott, Lister, Lupton, Mendoza~Magbanua, Malhotra, Mueller, Murtagh, Pandey, Reach, Samarasinha, Seligman, Snodgrass, Solontoi, Szabó, White, Womack, Young, Allbery, Armellin, Aubourg, Avdellidou, Azfar, Bauer, Bechtol, Belyakov, Benecchi, Bertini, Bolin, Bose, Buchanan, Boucaud, Boufleur, Boutigny, Braga-Ribas, Calabrese, Camargo, Caplar, Carry, Carvajal, Choi, Cowan, Croft, Ćuk, Daruich, Daubard, Davenport, Daylan, Delgado, Devillepoix, Doherty, Donaldson, Drass, Deppe, Dubois-Felsmann, Economou, Eduardo, Farnocchia, Frissell, Fedorets, Fernandes, Fulle, Gerdes, Gibbs, Gillan, Guy, Hammergren,
  Hanushevsky, Hernandez, Hestroffer, Hopkins, Granvik, Ieva, Irving, Jannuzi, Jimenez, Ramos Gomes-Júnior, Juramy, Kahn, Kannawadi, Kang, Kryszczyńska, Kotov, Koumjian, Krughoff, Lage, Lange, Levine, Li, Licandro, Lin, Lust, Lyttle, Mahabal, Mahlke, Plazas~Malagón, Salazar~Manzano, Marc, Margoti, Marčeta, Menanteau, Meyers, Mills, Morato, More, Morrison, Moulane, Muñoz-Gutiérrez, Newcomer~F., O'Connor, Oldag, Oldroyd, O'Mullane, Opitom, Oszkiewicz, Page, Patterson, Payne, Peloton, Pereira, Peterson, Polin, Pollek, Polen, Qiu, Ragozzine, Rajagopal, van Reeven, Rice, Ridgway, Rivkin, Robinson, Rożek, Salnikov, Sánchez, Sarid, Schambeau, Scolnic, Schindler, Seaman, Jacques, Shaw, Shugart, Sick, Siraj, Sitarz, Sobhani, Soldahl, Stalder, Stetzler, Swinbank, Szigeti, Tauraso, Thornton, Tonietti, Trilling, \& Trujillo}]{Chandler2025}
Chandler, C.~O., Bernardinelli, P.~H., Jurić, M., {et~al.} 2025, \bibinfo{title}{{NSF}-{DOE} {Vera} {C}. {Rubin} {Observatory} {Observations} of {Interstellar} {Comet} {3I}/{ATLAS} ({C}/2025 {N1}),} arXiv, \dodoi{10.48550/arXiv.2507.13409}

\bibitem[{A.~L. Cochran {et~al.}(2015)Cochran, Levasseur-Regourd, Cordiner, Hadamcik, Lasue, Gicquel, Schleicher, Charnley, Mumma, Paganini, Bockelée-Morvan, Biver, \& Kuan}]{Cochran2015}
Cochran, A.~L., Levasseur-Regourd, A.-C., Cordiner, M., {et~al.} 2015, \bibinfo{title}{The {Composition} of {Comets},} Space Science Reviews, 197, 9, \dodoi{10.1007/s11214-015-0183-6}

\bibitem[{W.~D. Cochran {et~al.}(2007)Cochran, Endl, Wittenmyer, \& Bean}]{Cochran2007}
Cochran, W.~D., Endl, M., Wittenmyer, R.~A., \& Bean, J.~L. 2007, \bibinfo{title}{A {Planetary} {System} around {HD} 155358: {The} {Lowest} {Metallicity} {Planet} {Host} {Star}*,} The Astrophysical Journal, 665, 1407, \dodoi{10.1086/519555}

\bibitem[{M.~A. Cordiner {et~al.}(2020)Cordiner, Milam, Biver, Bockelée-Morvan, Roth, Bergin, Jehin, Remijan, Charnley, Mumma, Boissier, Crovisier, Paganini, Kuan, \& Lis}]{Cordiner2020}
Cordiner, M.~A., Milam, S.~N., Biver, N., {et~al.} 2020, \bibinfo{title}{Unusually high {CO} abundance of the first active interstellar comet,} Nature Astronomy, 4, 861, \dodoi{10.1038/s41550-020-1087-2}

\bibitem[{R. de~la Fuente~Marcos {et~al.}(2025)de~la Fuente~Marcos, Alarcon, Licandro, Serra-Ricart, de~Leon, de~la Fuente~Marcos, Lombardi, Tejero, Cabrera-Lavers, Guerra~Arencibia, \& Ruiz~Cejudo}]{delaFuenteMarcos2025}
de~la Fuente~Marcos, R., Alarcon, M.~R., Licandro, J., {et~al.} 2025, \bibinfo{title}{Assessing interstellar comet {3I}/{ATLAS} with the 10.4 m {Gran} {Telescopio} {Canarias} and the {Two}-meter {Twin} {Telescope},} arXiv, \dodoi{10.48550/arXiv.2507.12922}

\bibitem[{W. Dehnen(1998)Dehnen}]{Dehnen1998}
Dehnen, W. 1998, \bibinfo{title}{The {Distribution} of {Nearby} {Stars} in {Velocity} {Space} {Inferred} from {HIPPARCOS} {Data},} The Astronomical Journal, 115, 2384, \dodoi{10.1086/300364}

\bibitem[{F.~E. DeMeo {et~al.}(2009)DeMeo, Binzel, Slivan, \& Bus}]{DeMeo2009}
DeMeo, F.~E., Binzel, R.~P., Slivan, S.~M., \& Bus, S.~J. 2009, \bibinfo{title}{An extension of the {Bus} asteroid taxonomy into the near-infrared,} Icarus, 202, 160, \dodoi{10.1016/j.icarus.2009.02.005}

\bibitem[{L. Denneau {et~al.}(2025)Denneau, Siverd, Tonry, Weiland, Erasmus, Fitzsimmons, \& Robinson}]{Denneau2025}
Denneau, L., Siverd, R., Tonry, J., {et~al.} 2025, \bibinfo{title}{{3I}/{ATLAS} = {C}/2025 {N1} ({ATLAS}),} MPEC

\bibitem[{S.~J. Desch \& A.~P. Jackson(2021)Desch \& Jackson}]{Desch2021}
Desch, S.~J., \& Jackson, A.~P. 2021, \bibinfo{title}{{1I}/`oumuamua as an {N}₂ ice fragment of an exo pluto surface {II}: {Generation} of {N}₂ ice fragments and the origin of `oumuamua,} Journal of Geophysical Research (Planets), 126, e06807, \dodoi{10.1029/2020JE006807}

\bibitem[{S.~J. Desch \& A.~P. Jackson(2022)Desch \& Jackson}]{Desch2022}
Desch, S.~J., \& Jackson, A.~P. 2022, \bibinfo{title}{Some pertinent issues for interstellar panspermia raised after the discovery of {1I}/`oumuamua,} Astrobiology, 22, 1400, \dodoi{10.1089/ast.2021.0199}

\bibitem[{J. de León {et~al.}(2020)de León, Licandro, de la Fuente Marcos, de la Fuente Marcos, Lara, Moreno, Pinilla-Alonso, Serra-Ricart, De Prá, Tozzi, Souza-Feliciano, Popescu, Scarpa, Font Serra, Geier, Lorenzi, Harutyunyan, \& Cabrera-Lavers}]{deLeon2020}
de León, J., Licandro, J., de la Fuente Marcos, C., {et~al.} 2020, \bibinfo{title}{Visible and near-infrared observations of interstellar comet {2I}/{Borisov} with the 10.4-m {GTC} and the 3.6-m {TNG} telescopes,} Monthly Notices of the Royal Astronomical Society, 495, 2053, \dodoi{10.1093/mnras/staa1190}

\bibitem[{A. Do {et~al.}(2018)Do, Tucker, \& Tonry}]{Do2018}
Do, A., Tucker, M.~A., \& Tonry, J. 2018, \bibinfo{title}{Interstellar {Interlopers}: {Number} {Density} and {Origin} of ‘{Oumuamua}-like {Objects},} The Astrophysical Journal Letters, 855, L10, \dodoi{10.3847/2041-8213/aaae67}

\bibitem[{R.~C. Dorsey {et~al.}(2025)Dorsey, Hopkins, Bannister, Lawler, Lintott, Parker, \& Forbes}]{Dorsey2025}
Dorsey, R.~C., Hopkins, M.~J., Bannister, M.~T., {et~al.} 2025, \bibinfo{title}{The visibility of the Ōtautahi-{Oxford} interstellar object population model in {LSST},} arXiv, \dodoi{10.48550/arXiv.2502.16741}

\bibitem[{M. Drahus {et~al.}(2017)Drahus, Yang, Lis, \& Jewitt}]{Drahus2017}
Drahus, M., Yang, B., Lis, D.~C., \& Jewitt, D. 2017, \bibinfo{title}{New limits to {CO} outgassing in {Centaurs},} Monthly Notices of the Royal Astronomical Society, 468, 2897, \dodoi{10.1093/mnras/stw2227}

\bibitem[{A.~D. Feinstein {et~al.}(2025)Feinstein, Noonan, \& Seligman}]{Feinstein2025}
Feinstein, A.~D., Noonan, J.~W., \& Seligman, D.~Z. 2025, \bibinfo{title}{Precovery {Observations} of {3I}/{ATLAS} from {TESS} {Suggests} {Possible} {Distant} {Activity},} arXiv, \dodoi{10.48550/arXiv.2507.21967}

\bibitem[{F. Feng \& H.~R.~A. Jones(2018)Feng \& Jones}]{Feng2018}
Feng, F., \& Jones, H. R.~A. 2018, \bibinfo{title}{‘{Oumuamua} as a {Messenger} from the {Local} {Association},} The Astrophysical Journal Letters, 852, L27, \dodoi{10.3847/2041-8213/aaa404}

\bibitem[{A. Fitzsimmons {et~al.}(2023)Fitzsimmons, Meech, Matrà, \& Pfalzner}]{Fitzsimmons2023}
Fitzsimmons, A., Meech, K., Matrà, L., \& Pfalzner, S. 2023, \bibinfo{title}{Interstellar objects and exocomets,} arXiv e-prints, arXiv:2303.17980, \dodoi{10.48550/arXiv.2303.17980}

\bibitem[{E.~G. Flekkøy {et~al.}(2019)Flekkøy, Luu, \& Toussaint}]{Flekkoy2019}
Flekkøy, E.~G., Luu, J., \& Toussaint, R. 2019, \bibinfo{title}{The {Interstellar} {Object} ’{Oumuamua} as a {Fractal} {Dust} {Aggregate},} The Astrophysical Journal Letters, 885, L41, \dodoi{10.3847/2041-8213/ab4f78}

\bibitem[{J.~C. Forbes {et~al.}(2025)Forbes, Bannister, Lintott, Forrest, Portegies~Zwart, Dorsey, Albrow, \& Hopkins}]{Forbes2025}
Forbes, J.~C., Bannister, M.~T., Lintott, C., {et~al.} 2025, \bibinfo{title}{He {Awa} {Whiria}: {The} {Tidal} {Streams} of {Interstellar} {Objects},} The Astrophysical Journal, 988, 121, \dodoi{10.3847/1538-4357/adc9ac}

\bibitem[{J.~C. Forbes \& A. Loeb(2019)Forbes \& Loeb}]{Forbes2019}
Forbes, J.~C., \& Loeb, A. 2019, \bibinfo{title}{Turning up the {Heat} on ‘{Oumuamua},} The Astrophysical Journal, 875, L23, \dodoi{10.3847/2041-8213/ab158f}

\bibitem[{W.~C. Fraser {et~al.}(2018)Fraser, Pravec, Fitzsimmons, Lacerda, Bannister, Snodgrass, \& Smolić}]{Fraser2018}
Fraser, W.~C., Pravec, P., Fitzsimmons, A., {et~al.} 2018, \bibinfo{title}{The tumbling rotational state of {1I}/`{Oumuamua},} Nature Astronomy, 2, 383, \dodoi{10.1038/s41550-018-0398-z}

\bibitem[{E. Gaidos {et~al.}(2017)Gaidos, Williams, \& Kraus}]{Gaidos2017}
Gaidos, E., Williams, J., \& Kraus, A. 2017, \bibinfo{title}{Origin of interstellar object a/2017 {U1} in a nearby young stellar association?} RNAAS, 1, 13, \dodoi{10.3847/2515-5172/aa9851}

\bibitem[{T. Hallatt \& P. Wiegert(2020)Hallatt \& Wiegert}]{Hallatt2020}
Hallatt, T., \& Wiegert, P. 2020, \bibinfo{title}{The {Dynamics} of {Interstellar} {Asteroids} and {Comets} within the {Galaxy}: {An} {Assessment} of {Local} {Candidate} {Source} {Regions} for {1I}/’{Oumuamua} and {2I}/{Borisov},} The Astronomical Journal, 159, 147, \dodoi{10.3847/1538-3881/ab7336}

\bibitem[{J. Holmberg {et~al.}(2009)Holmberg, Nordström, \& Andersen}]{Holmberg2009}
Holmberg, J., Nordström, B., \& Andersen, J. 2009, \bibinfo{title}{The {Geneva}-{Copenhagen} survey of the solar neighbourhood. {III}. {Improved} distances, ages, and kinematics,} Astronomy and Astrophysics, 501, 941, \dodoi{10.1051/0004-6361/200811191}

\bibitem[{D.~J. Hoover {et~al.}(2022)Hoover, Seligman, \& Payne}]{Hoover2022}
Hoover, D.~J., Seligman, D.~Z., \& Payne, M.~J. 2022, \bibinfo{title}{The {Population} of {Interstellar} {Objects} {Detectable} with the {LSST} and {Accessible} for {In} {Situ} {Rendezvous} with {Various} {Mission} {Designs},} The Planetary Science Journal, 3, 71, \dodoi{10.3847/PSJ/ac58fe}

\bibitem[{M.~J. Hopkins {et~al.}(2025{\natexlab{a}})Hopkins, Bannister, \& Lintott}]{Hopkins2025a}
Hopkins, M.~J., Bannister, M.~T., \& Lintott, C. 2025{\natexlab{a}}, \bibinfo{title}{Predicting {Interstellar} {Object} {Chemodynamics} with {Gaia},} The Astronomical Journal, 169, 78, \dodoi{10.3847/1538-3881/ad9eb3}

\bibitem[{M.~J. Hopkins {et~al.}(2025{\natexlab{b}})Hopkins, Dorsey, Forbes, Bannister, Lintott, \& Leicester}]{Hopkins2025}
Hopkins, M.~J., Dorsey, R.~C., Forbes, J.~C., {et~al.} 2025{\natexlab{b}}, \bibinfo{title}{From a {Different} {Star}: {3I}/{ATLAS} in the context of the Ōtautahi-{Oxford} interstellar object population model,} arXiv, \dodoi{10.48550/arXiv.2507.05318}

\bibitem[{M.~J. Hopkins {et~al.}(2023)Hopkins, Lintott, Bannister, Mackereth, \& Forbes}]{Hopkins2023}
Hopkins, M.~J., Lintott, C., Bannister, M.~T., Mackereth, J.~T., \& Forbes, J.~C. 2023, \bibinfo{title}{The {Galactic} {Interstellar} {Object} {Population}: {A} {Framework} for {Prediction} and {Inference},} The Astronomical Journal, 166, 241, \dodoi{10.3847/1538-3881/ad03e6}

\bibitem[{C.-H. Hsieh {et~al.}(2021)Hsieh, Laughlin, \& Arce}]{Hsieh2021}
Hsieh, C.-H., Laughlin, G., \& Arce, H.~G. 2021, \bibinfo{title}{Evidence {Suggesting} {That} ‘{Oumuamua} {Is} the ∼30 {Myr} {Old} {Product} of a {Molecular} {Cloud},} The Astrophysical Journal, 917, 20, \dodoi{10.3847/1538-4357/ac0729}

\bibitem[{A.~P. Jackson \& S.~J. Desch(2021)Jackson \& Desch}]{Jackson2021}
Jackson, A.~P., \& Desch, S.~J. 2021, \bibinfo{title}{{1I}/`oumuamua as an {N}₂ ice fragment of an exo pluto surface: {I}. {Size} and compositional constraints,} Journal of Geophysical Research (Planets), 126, e06706, \dodoi{10.1029/2020JE006706}

\bibitem[{D. Jewitt {et~al.}(2025)Jewitt, Hui, Mutchler, Kim, \& Agarwal}]{Jewitt2025a}
Jewitt, D., Hui, M.-T., Mutchler, M., Kim, Y., \& Agarwal, J. 2025, \bibinfo{title}{Hubble {Space} {Telescope} {Observations} of the {Interstellar} {Interloper} {3I}/{ATLAS},} arXiv.
\newblock \url{https://ui.adsabs.harvard.edu/abs/2025arXiv250802934J}

\bibitem[{D. Jewitt {et~al.}(2020)Jewitt, Kim, Mutchler, Weaver, Agarwal, \& Hui}]{Jewitt2020}
Jewitt, D., Kim, Y., Mutchler, M., {et~al.} 2020, \bibinfo{title}{Outburst and {Splitting} of {Interstellar} {Comet} {2I}/{Borisov},} The Astrophysical Journal Letters, 896, L39, \dodoi{10.3847/2041-8213/ab99cb}

\bibitem[{D. Jewitt \& J. Luu(2019{\natexlab{a}})Jewitt \& Luu}]{Jewitt2019a}
Jewitt, D., \& Luu, J. 2019{\natexlab{a}}, \bibinfo{title}{Initial {Characterization} of {Interstellar} {Comet} {2I}/2019 {Q4} ({Borisov}),} The Astrophysical Journal Letters, 886, L29, \dodoi{10.3847/2041-8213/ab530b}

\bibitem[{D. Jewitt \& J. Luu(2019{\natexlab{b}})Jewitt \& Luu}]{Jewitt2019}
Jewitt, D., \& Luu, J. 2019{\natexlab{b}}, \bibinfo{title}{Disintegrating {Inbound} {Long}-period {Comet} {C}/2019 {J2},} The Astrophysical Journal Letters, 883, L28, \dodoi{10.3847/2041-8213/ab4135}

\bibitem[{D. Jewitt \& J. Luu(2025)Jewitt \& Luu}]{Jewitt2025}
Jewitt, D., \& Luu, J. 2025, \bibinfo{title}{Interstellar {Interloper} {C}/2025 {N1} is {Active},} \{The Astronomer's Telegram\} No. 17263

\bibitem[{D. Jewitt {et~al.}(2017)Jewitt, Luu, Rajagopal, Kotulla, Ridgway, Liu, \& Augusteijn}]{Jewitt2017}
Jewitt, D., Luu, J., Rajagopal, J., {et~al.} 2017, \bibinfo{title}{Interstellar interloper {1I}/2017 {U1}: {Observations} from the {NOT} and {WIYN} telescopes,} ApJL, 850, L36, \dodoi{10.3847/2041-8213/aa9b2f}

\bibitem[{D. Jewitt \& D.~Z. Seligman(2023)Jewitt \& Seligman}]{Jewitt2023}
Jewitt, D., \& Seligman, D.~Z. 2023, \bibinfo{title}{The {Interstellar} {Interlopers},} Annual Review of Astronomy and Astrophysics, 61, 197, \dodoi{10.1146/annurev-astro-071221-054221}

\bibitem[{T. Kareta {et~al.}(2020)Kareta, Andrews, Noonan, Harris, Smith, O’Brien, Sharkey, Reddy, Springmann, Lejoly, Volk, Conrad, \& Veillet}]{Kareta2020}
Kareta, T., Andrews, J., Noonan, J.~W., {et~al.} 2020, \bibinfo{title}{Carbon {Chain} {Depletion} of {2I}/{Borisov},} The Astrophysical Journal Letters, 889, L38, \dodoi{10.3847/2041-8213/ab6a08}

\bibitem[{W.~G. Levine \& G. Laughlin(2021)Levine \& Laughlin}]{Levine2021}
Levine, W.~G., \& Laughlin, G. 2021, \bibinfo{title}{Assessing the {Formation} of {Solid} {Hydrogen} {Objects} in {Starless} {Molecular} {Cloud} {Cores},} The Astrophysical Journal, 912, 3, \dodoi{10.3847/1538-4357/abec85}

\bibitem[{W.~G. Levine {et~al.}(2023)Levine, Taylor, Seligman, Hoover, Jedicke, Bergner, \& Laughlin}]{Levine2023}
Levine, W.~G., Taylor, A.~G., Seligman, D.~Z., {et~al.} 2023, \bibinfo{title}{Interstellar {Comets} from {Post}-main-sequence {Systems} as {Tracers} of {Extrasolar} {Oort} {Clouds},} The Planetary Science Journal, 4, 124, \dodoi{10.3847/PSJ/acdf58}

\bibitem[{H.~W. Lin {et~al.}(2020)Lin, Lee, Gerdes, Adams, Becker, Napier, \& Markwardt}]{Lin2020}
Lin, H.~W., Lee, C.-H., Gerdes, D.~W., {et~al.} 2020, \bibinfo{title}{Detection of {Diatomic} {Carbon} in {2I}/{Borisov},} The Astrophysical Journal Letters, 889, L30, \dodoi{10.3847/2041-8213/ab6bd9}

\bibitem[{C. Lintott {et~al.}(2022)Lintott, Bannister, \& Mackereth}]{Lintott2022}
Lintott, C., Bannister, M.~T., \& Mackereth, J.~T. 2022, \bibinfo{title}{Predicting the {Water} {Content} of {Interstellar} {Objects} from {Galactic} {Star} {Formation} {Histories},} The Astrophysical Journal, 924, L1, \dodoi{10.3847/2041-8213/ac41d5}

\bibitem[{A. Loeb(2025)Loeb}]{Loeb2025}
Loeb, A. 2025, \bibinfo{title}{Comment on "{Discovery} and {Preliminary} {Characterization} of a {Third} {Interstellar} {Object}: {3I}/{ATLAS}" [{arXiv}:2507.02757],} arXiv, \dodoi{10.48550/arXiv.2507.05881}

\bibitem[{E. Mamajek(2017)Mamajek}]{Mamajek2017}
Mamajek, E. 2017, \bibinfo{title}{Kinematics of the {Interstellar} {Vagabond} {1I}/‘{Oumuamua} ({A}/2017 {U1}),} Research Notes of the American Astronomical Society, 1, 21, \dodoi{10.3847/2515-5172/aa9bdc}

\bibitem[{V.~A. Marsakov {et~al.}(2011)Marsakov, Koval', Borkova, \& Shapovalov}]{Marsakov2011}
Marsakov, V.~A., Koval', V.~V., Borkova, T.~V., \& Shapovalov, M.~V. 2011, \bibinfo{title}{The age-metallicity relation in the thin disk of the galaxy,} Astronomy Reports, 55, 667, \dodoi{10.1134/S1063772911080063}

\bibitem[{D. Marčeta \& D.~Z. Seligman(2023)Marčeta \& Seligman}]{Marceta2023}
Marčeta, D., \& Seligman, D.~Z. 2023, \bibinfo{title}{Synthetic {Detections} of {Interstellar} {Objects} with the {Rubin} {Observatory} {Legacy} {Survey} of {Space} and {Time},} The Planetary Science Journal, 4, 230, \dodoi{10.3847/PSJ/ad08c1}

\bibitem[{S. Mashchenko(2019)Mashchenko}]{Mashchenko2019}
Mashchenko, S. 2019, \bibinfo{title}{Modelling the light curve of `{Oumuamua}: evidence for torque and disc-like shape,} Monthly Notices of the Royal Astronomical Society, 489, 3003, \dodoi{10.1093/mnras/stz2380}

\bibitem[{F.~J. Massey(1951)Massey}]{Massey1951}
Massey, F.~J. 1951, \bibinfo{title}{The {Kolmogorov}-{Smirnov} {Test} for {Goodness} of {Fit},} Journal of the American Statistical Association, 46, 68, \dodoi{10.2307/2280095}

\bibitem[{K.~J. Meech {et~al.}(2017)Meech, Weryk, Micheli, Kleyna, Hainaut, Jedicke, Wainscoat, Chambers, Keane, Petric, Denneau, Magnier, Berger, Huber, Flewelling, Waters, Schunova-Lilly, \& Chastel}]{Meech2017}
Meech, K.~J., Weryk, R., Micheli, M., {et~al.} 2017, \bibinfo{title}{A brief visit from a red and extremely elongated interstellar asteroid,} Nature, 552, 378, \dodoi{10.1038/nature25020}

\bibitem[{M. Micheli {et~al.}(2018)Micheli, Farnocchia, Meech, Buie, Hainaut, Prialnik, Schörghofer, Weaver, Chodas, Kleyna, Weryk, Wainscoat, Ebeling, Keane, Chambers, Koschny, \& Petropoulos}]{Micheli2018}
Micheli, M., Farnocchia, D., Meech, K.~J., {et~al.} 2018, \bibinfo{title}{Non-gravitational acceleration in the trajectory of {1I}/2017 {U1} (‘{Oumuamua}),} Nature, 559, 223, \dodoi{10.1038/s41586-018-0254-4}

\bibitem[{A. Moro-Mart\'in(2022)Moro-Mart\'in}]{Moro-Martin2022a}
Moro-Mart\'in, A. 2022, \bibinfo{title}{Interstellar planetesimals,} arXiv e-prints, arXiv:2205.04277, \dodoi{10.48550/arXiv.2205.04277}

\bibitem[{A. Moro-Martín(2019)Moro-Martín}]{Moro-Martin2019}
Moro-Martín, A. 2019, \bibinfo{title}{Could {1I}/’{Oumuamua} be an {Icy} {Fractal} {Aggregate}?} The Astrophysical Journal, 872, L32, \dodoi{10.3847/2041-8213/ab05df}

\bibitem[{ {MPEC}(2017){MPEC}}]{MPEC2017}
{MPEC}. 2017, \bibinfo{title}{The international astronomical union minor planet center,} IAU Minor Planet Center.
\newblock \url{https://minorplanetcenter.net//mpec/K17/K17UI1.html}

\bibitem[{B. Nordström {et~al.}(2004)Nordström, Mayor, Andersen, Holmberg, Pont, Jørgensen, Olsen, Udry, \& Mowlavi}]{Nordstrom2004}
Nordström, B., Mayor, M., Andersen, J., {et~al.} 2004, \bibinfo{title}{The {Geneva}-{Copenhagen} survey of the {Solar} neighbourhood. {Ages}, metallicities, and kinematic properties of ∼14 000 {F} and {G} dwarfs,} Astronomy and Astrophysics, 418, 989, \dodoi{10.1051/0004-6361:20035959}

\bibitem[{C. Opitom {et~al.}(2025)Opitom, Snodgrass, Jehin, Bannister, Bufanda, Deam, Dorsey, Ferrais, Hmiddouch, Knight, Kokotanekova, Leicester, Marsset, Murphy, Okoth, Ridden-Harper, Vander~Donckt, Ferellec, Hutsemekers, Lippi, Manfroid, \& Benkhaldoun}]{Opitom2025}
Opitom, C., Snodgrass, C., Jehin, E., {et~al.} 2025, \bibinfo{title}{Snapshot of a new interstellar comet: {3I}/{ATLAS} has a red and featureless spectrum,} arXiv, \dodoi{10.48550/arXiv.2507.05226}

\bibitem[{R.~S. Park {et~al.}(2018)Park, Pisano, Lazio, Chodas, \& Naidu}]{Park2018}
Park, R.~S., Pisano, D.~J., Lazio, T. J.~W., Chodas, P.~W., \& Naidu, S.~P. 2018, \bibinfo{title}{Search for {OH} 18 cm {Radio} {Emission} from {1I}/2017 {U1} with the {Green} {Bank} {Telescope},} The Astronomical Journal, 155, 185, \dodoi{10.3847/1538-3881/aab78d}

\bibitem[{S. Portegies~Zwart(2021)Portegies~Zwart}]{PortegiesZwart2021}
Portegies~Zwart, S. 2021, \bibinfo{title}{Oort cloud {Ecology} - {I}. {Extra}-solar {Oort} clouds and the origin of asteroidal interlopers,} Astronomy \& Astrophysics, 647, A136, \dodoi{10.1051/0004-6361/202038888}

\bibitem[{S. Portegies Zwart {et~al.}(2018)Portegies Zwart, Torres, Pelupessy, Bédorf, \& Cai}]{PortegiesZwart2018}
Portegies Zwart, S., Torres, S., Pelupessy, I., Bédorf, J., \& Cai, M.~X. 2018, \bibinfo{title}{The origin of interstellar asteroidal objects like {1I}/2017 {U1} ‘{Oumuamua},} Monthly Notices of the Royal Astronomical Society: Letters, 479, L17, \dodoi{10.1093/mnrasl/sly088}

\bibitem[{T.~H. Puzia {et~al.}(2025)Puzia, Rahatgaonkar, Carvajal, Nayak, \& Luco}]{Puzia2025}
Puzia, T.~H., Rahatgaonkar, R., Carvajal, J.~P., Nayak, P.~K., \& Luco, B. 2025, \bibinfo{title}{Spectral {Characteristics} of {Interstellar} {Object} {3I}/{ATLAS} from {SOAR} {Observations},} arXiv.
\newblock \url{https://ui.adsabs.harvard.edu/abs/2025arXiv250802777P}

\bibitem[{P. Ramos {et~al.}(2018)Ramos, Antoja, \& Figueras}]{Ramos2018}
Ramos, P., Antoja, T., \& Figueras, F. 2018, \bibinfo{title}{Riding the kinematic waves in the {Milky} {Way} disk with {Gaia},} Astronomy and Astrophysics, 619, A72, \dodoi{10.1051/0004-6361/201833494}

\bibitem[{T. Santana-Ros {et~al.}(2025)Santana-Ros, Ivanova, Mykhailova, Erasmus, Kamiński, Oszkiewicz, Kwiatkowski, Husárik, Ngwane, \& Penttilä}]{Santana-Ros2025}
Santana-Ros, T., Ivanova, O., Mykhailova, S., {et~al.} 2025, \bibinfo{title}{Temporal {Evolution} of the {Third} {Interstellar} {Comet} {3I}/{ATLAS}: {Spin}, {Color}, {Spectra} and {Dust} {Activity},} arXiv, \dodoi{10.48550/arXiv.2508.00808}

\bibitem[{Z. Sekanina(2019)Sekanina}]{Sekanina2019}
Sekanina, Z. 2019, \bibinfo{title}{{1I}/`oumuamua as debris of dwarf interstellar comet that disintegrated before perihelion,} arXiv e-prints, arXiv:1901.08704, \dodoi{10.48550/arXiv.1901.08704}

\bibitem[{D. Seligman \& G. Laughlin(2018)Seligman \& Laughlin}]{Seligman2018}
Seligman, D., \& Laughlin, G. 2018, \bibinfo{title}{The {Feasibility} and {Benefits} of {In} {Situ} {Exploration} of ‘{Oumuamua}-like {Objects},} The Astronomical Journal, 155, 217, \dodoi{10.3847/1538-3881/aabd37}

\bibitem[{D. Seligman \& G. Laughlin(2020)Seligman \& Laughlin}]{Seligman2020}
Seligman, D., \& Laughlin, G. 2020, \bibinfo{title}{Evidence that {1I}/2017 {U1} (‘{Oumuamua}) was {Composed} of {Molecular} {Hydrogen} {Ice},} The Astrophysical Journal Letters, 896, L8, \dodoi{10.3847/2041-8213/ab963f}

\bibitem[{D.~Z. Seligman \& A. Moro-Mart\'in(2023)Seligman \& Moro-Mart\'in}]{Seligman2023a}
Seligman, D.~Z., \& Moro-Mart\'in, A. 2023, \bibinfo{title}{Interstellar objects,} Contemporary Physics, 63, 200, \dodoi{10.1080/00107514.2023.2203976}

\bibitem[{D.~Z. Seligman {et~al.}(2025)Seligman, Micheli, Farnocchia, Denneau, Noonan, Santana-Ros, Conversi, Devogèle, Faggioli, Feinstein, Fenucci, Frincke, Hainaut, Hoogendam, Hsieh, Kareta, Kelley, Lister, Marčeta, Meech, Ocaña, Peña-Asensio, Shappee, Taylor, Wainscoat, Weryk, Wray, Yaginuma, Yang, \& Ye}]{Seligman2025}
Seligman, D.~Z., Micheli, M., Farnocchia, D., {et~al.} 2025, \bibinfo{title}{Discovery and {Preliminary} {Characterization} of a {Third} {Interstellar} {Object}: {3I}/{ATLAS},} arXiv, \dodoi{10.48550/arXiv.2507.02757}

\bibitem[{O. Snaith {et~al.}(2015)Snaith, Haywood, Di~Matteo, Lehnert, Combes, Katz, \& Gómez}]{Snaith2015}
Snaith, O., Haywood, M., Di~Matteo, P., {et~al.} 2015, \bibinfo{title}{Reconstructing the star formation history of the {Milky} {Way} disc(s) from chemical abundances,} Astronomy and Astrophysics, 578, A87, \dodoi{10.1051/0004-6361/201424281}

\bibitem[{A.~G. Taylor {et~al.}(2023)Taylor, Seligman, Hainaut, \& Meech}]{Taylor2023a}
Taylor, A.~G., Seligman, D.~Z., Hainaut, O.~R., \& Meech, K.~J. 2023, \bibinfo{title}{Fitting the {Light} {Curve} of {1I}/‘{Oumuamua} with a {Nonprincipal} {Axis} {Rotational} {Model} and {Outgassing} {Torques},} The Planetary Science Journal, 4, 186, \dodoi{10.3847/PSJ/acf617}

\bibitem[{J.~L. Tonry {et~al.}(2018)Tonry, Denneau, Heinze, Stalder, Smith, Smartt, Stubbs, Weiland, \& Rest}]{Tonry2018}
Tonry, J.~L., Denneau, L., Heinze, A.~N., {et~al.} 2018, \bibinfo{title}{{ATLAS}: {A} {High}-cadence {All}-sky {Survey} {System},} Publications of the Astronomical Society of the Pacific, 130, 064505, \dodoi{10.1088/1538-3873/aabadf}

\bibitem[{D.~E. Trilling {et~al.}(2018)Trilling, Mommert, Hora, Farnocchia, Chodas, Giorgini, Smith, Carey, Lisse, Werner, McNeill, Chesley, Emery, Fazio, Fernandez, Harris, Marengo, Mueller, Roegge, Smith, Weaver, Meech, \& Micheli}]{Trilling2018}
Trilling, D.~E., Mommert, M., Hora, J.~L., {et~al.} 2018, \bibinfo{title}{Spitzer {Observations} of {Interstellar} {Object} {1I}/‘{Oumuamua},} The Astronomical Journal, 156, 261, \dodoi{10.3847/1538-3881/aae88f}

\bibitem[{G.~V. Williams {et~al.}(2017)Williams, Sato, Sarneczky, Wainscoat, Woodworth, \& Meech}]{Williams2017}
Williams, G.~V., Sato, H., Sarneczky, K., {et~al.} 2017, \bibinfo{title}{Minor planets 2017 {SN}\_33 and 2017 {U1},} Central Bureau Electronic Telegrams, 4450, 1

\bibitem[{M. Xiang \& H.-W. Rix(2022)Xiang \& Rix}]{Xiang2022}
Xiang, M., \& Rix, H.-W. 2022, \bibinfo{title}{A time-resolved picture of our {Milky} {Way}’s early formation history,} Nature, 603, 599, \dodoi{10.1038/s41586-022-04496-5}

\bibitem[{Z. Xing {et~al.}(2020)Xing, Bodewits, Noonan, \& Bannister}]{Xing2020}
Xing, Z., Bodewits, D., Noonan, J., \& Bannister, M.~T. 2020, \bibinfo{title}{Water {Production} {Rates} and {Activity} of {Interstellar} {Comet} {2I}/{Borisov},} The Astrophysical Journal Letters, 893, L48, \dodoi{10.3847/2041-8213/ab86be}

\bibitem[{A. Yaginuma {et~al.}(2025)Yaginuma, Frincke, Seligman, Mandt, DellaGiustina, Peña-Asensio, Taylor, \& Nolan}]{Yaginuma2025}
Yaginuma, A., Frincke, T., Seligman, D.~Z., {et~al.} 2025, \bibinfo{title}{The {Feasibility} of a {Spacecraft} {Flyby} with the {Third} {Interstellar} {Object} {3I}/{ATLAS} from {Earth} or {Mars},} arXiv, \dodoi{10.48550/arXiv.2507.15755}

\bibitem[{B. Yang {et~al.}(2025)Yang, Meech, Connelley, \& Keane}]{Yang2025}
Yang, B., Meech, K.~J., Connelley, M., \& Keane, J.~V. 2025, \bibinfo{title}{Spectroscopic {Characterization} of {Interstellar} {Object} {3I}/{ATLAS}: {Water} {Ice} in the {Coma},} arXiv, \dodoi{10.48550/arXiv.2507.14916}

\end{thebibliography}
\bibliographystyle{aasjournal}

\end{document}